\documentclass[12pt,a4paper]{article} 

\usepackage{graphicx,subfigure,amsmath,cite}
\renewcommand{\Re}{\textrm{Re}\,}
\newcommand{\eps}{\varepsilon}

\title{Light scattering by a subwavelength plasmonic array: anisotropic model}

\author{Anton Nemykin, Leonid Frumin, David Shapiro\\
	Institute of Automation and Electrometry\\
	Russian Academy of Sciences, Siberian Branch\\
	Novosibirsk 630090, 1 Koptjug Ave, Russia\\
e-mail: shapiro@iae.nsk.su}

\begin{document}
\maketitle

\begin{abstract}
	We calculate the light transmission by a subwavelength plasmonic array using the boundary element method for parallel cylinders with different cross-sections: circular or elliptic with axes ratio 4:1. We demonstrate that the plasmonic resonance is sharper for the case of ellipses with the large horizontal axis. This structure is susceptible to the refractive index variations of media since the high derivative of reflection and transmission coefficients are near the angle of total internal reflection. To obtain an approximate analytical expression, we use the model of a metallic layer. We explore the "sandwich" structure with an anisotropic film between two dielectrics and demonstrate its quantitative agreement with numerical results.
\end{abstract}

\section{Introduction}
Plasmons are surface waves of conduction electrons inside the metallic film near the border with a dielectric. The dispersion relation for plasmons differs from the bulk plasma due to an interaction with the dielectric medium. The plasmon electric field decays when one penetrates deeper both into the metal and dielectric areas. At plasmon resonance, there is a notable increase in the electric field, and intensity \cite{zayats2005nano}.

Applications of surface plasmons are diverse: from optical biosensors \cite{gupta2018recent,xu2019optical,zhao2019current} to the acceleration of relativistic electrons \cite{peralta2013demonstration} and space jet engines \cite{rovey2015plasmonic}. 
Let us consider the lower dielectric half-space as the substrate with a planar periodic structure on top. The refractive index of the upper half-space strongly affects the angular and wavelength spectral characteristics of the scattering layer due to crucial changes in layer effective permittivity. These resonance responses underlie plasmon biosensors. The scattering on the layer is also essential for research in silicon photonics, such as the study, development, and manufacture of optical microcircuits in which photons propagate instead of electrons \cite{bogaerts2018silicon}. The all-optical circuits can significantly enhance the density of communication channels and lead to considerable energy savings at once.

Recently the study of the plasmon-enhanced local field in the array of parallel circular metallic cylinders is proposed to improve the sensitivity to refractive index variations \cite{frumin2020sensitivity}. The matter is rapid changes in the Fresnel coefficients near the angle of total internal reflection (TIR). Earlier, we analyzed the grid of parallel metallic cylinders between two dielectrics; see \cite{frumin2013plasmons} and references therein. The boundary element method (BEM) idea is to transform the  Maxwell equations to the boundary integral ones applying the Green theorem. In the case of a grid with parallel cylinders, the Floquet theorem reduces the problem of infinite grating to that with one cylinder within one elementary cell. To get analytical formulas, we exploited simplified models. The sequence of nanowires can be approximately replaced by a thin layer with the averaged permittivity \cite{nemykin2015excitation}. In the present paper, we treat cylinders of the elliptic cross-section with different orientations and compare the results to more general models of an anisotropic layer.

\section{FEF calculation}
\begin{figure}\centering
\includegraphics[width=0.5\columnwidth]{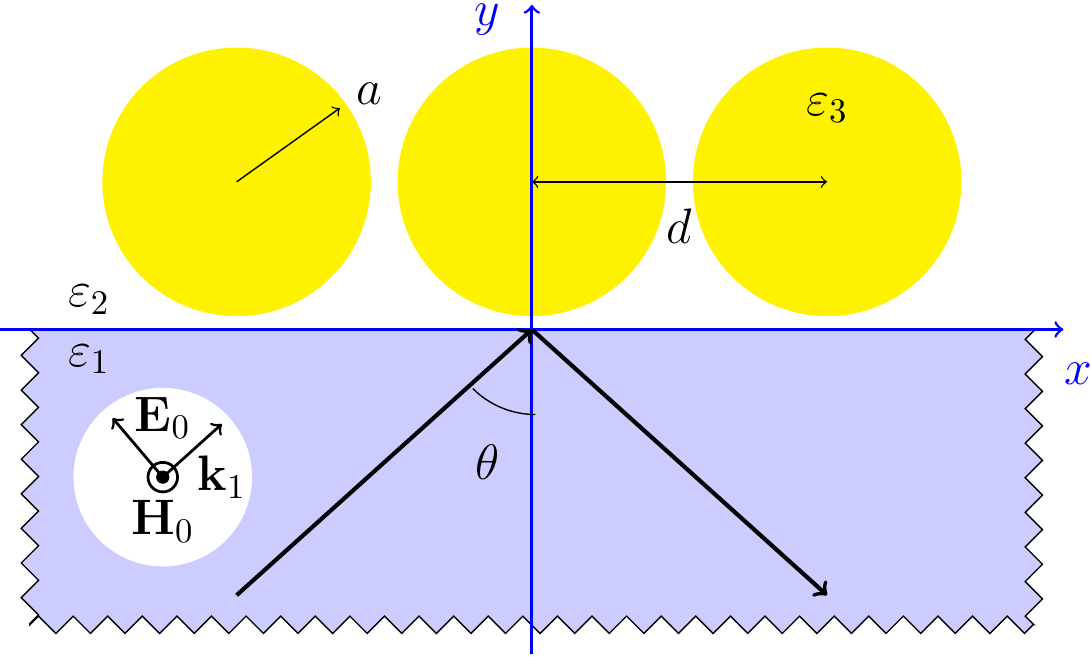}\includegraphics[width=0.9in]{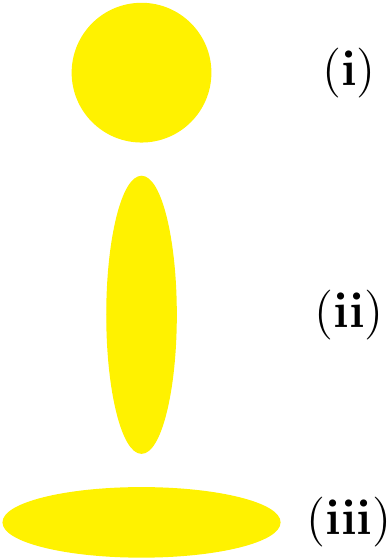}	
\caption{The periodic set of circular cylinders near the interface between free space and the glass. The inset describes the shapes of cross-section considered: circle (i), vertical ellipse (ii), horizontal (iii).}	
\label{f:sketch}
\end{figure}

We carry out the numerical modeling by the BEM \cite{frumin2013plasmons} for cylinders near the interface between upper free half-space ($\eps_2=1$) and glass lower half-space $\eps_1=2.25$.  We analyze an infinite grating with period $d$ consisting of gold ($\eps_3=-23.6+i1.27$ at $\lambda=0.7749~\mu$m \cite{palik98}). We consider three shapes of cylinder cross-sections: circle of radius $a=0.05~\mu$m and ellipses with axial ratio 1:4 and 4:1, as shown by an inset in fig. \ref{f:sketch}. The cross-section area is fixed, and the gap dimension is $0.01~\mu$m for all the cases; meanwhile, the period is different. It is $d=0.11~\mu$m for circular cylinders, $0.06~\mu$m for vertical ellipse, and $0.21~\mu$m for horizontal.  The distance between their centers and glass half-space is fixed $0.11~\mu$m for all the cases. The incidence field is $p$-wave near angle $\theta=\theta_0=41.81^\circ$, where $\theta_0$ is the TIR angle. The grid sampling in calculations is nonuniform near $\theta_0$ by a power law with exponents 3/2. 

\begin{figure}\centering
	\subfigure[]{\includegraphics[width=0.7\columnwidth]{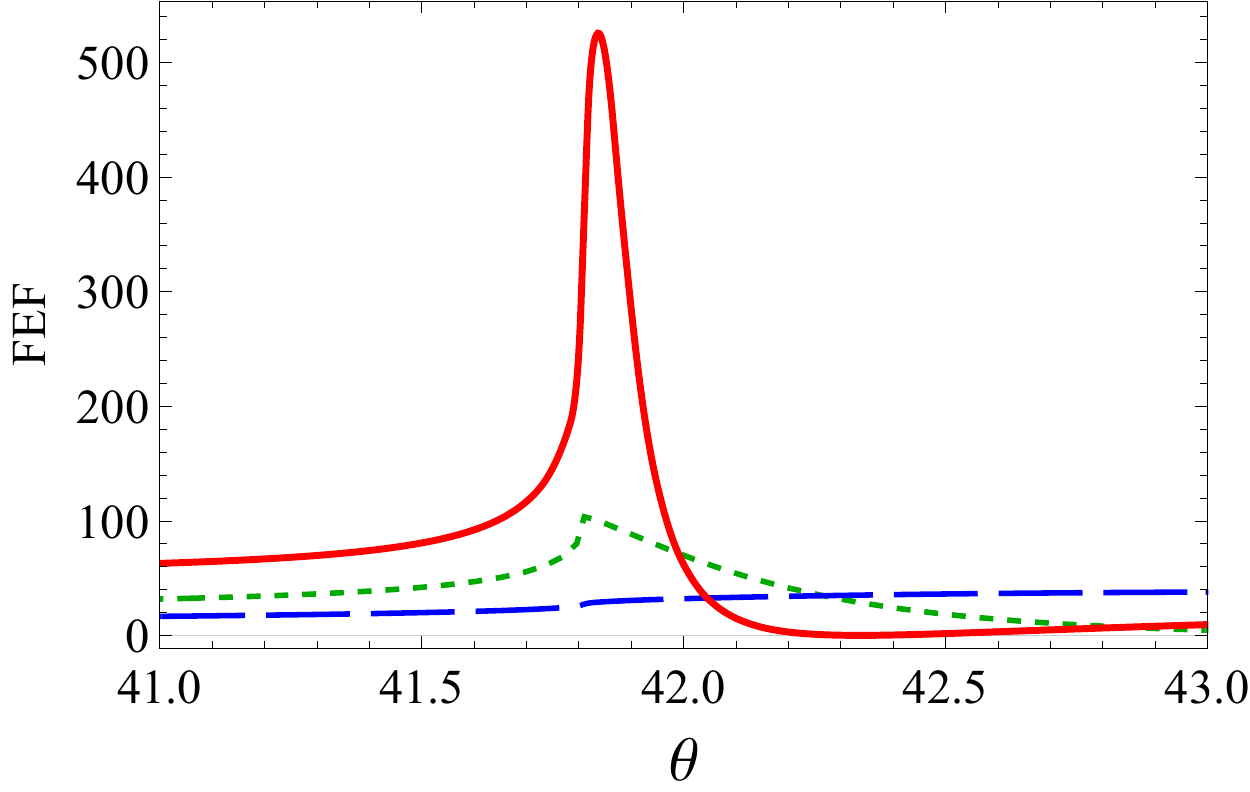}}
	
	\subfigure[]{\includegraphics[width=0.7\columnwidth]{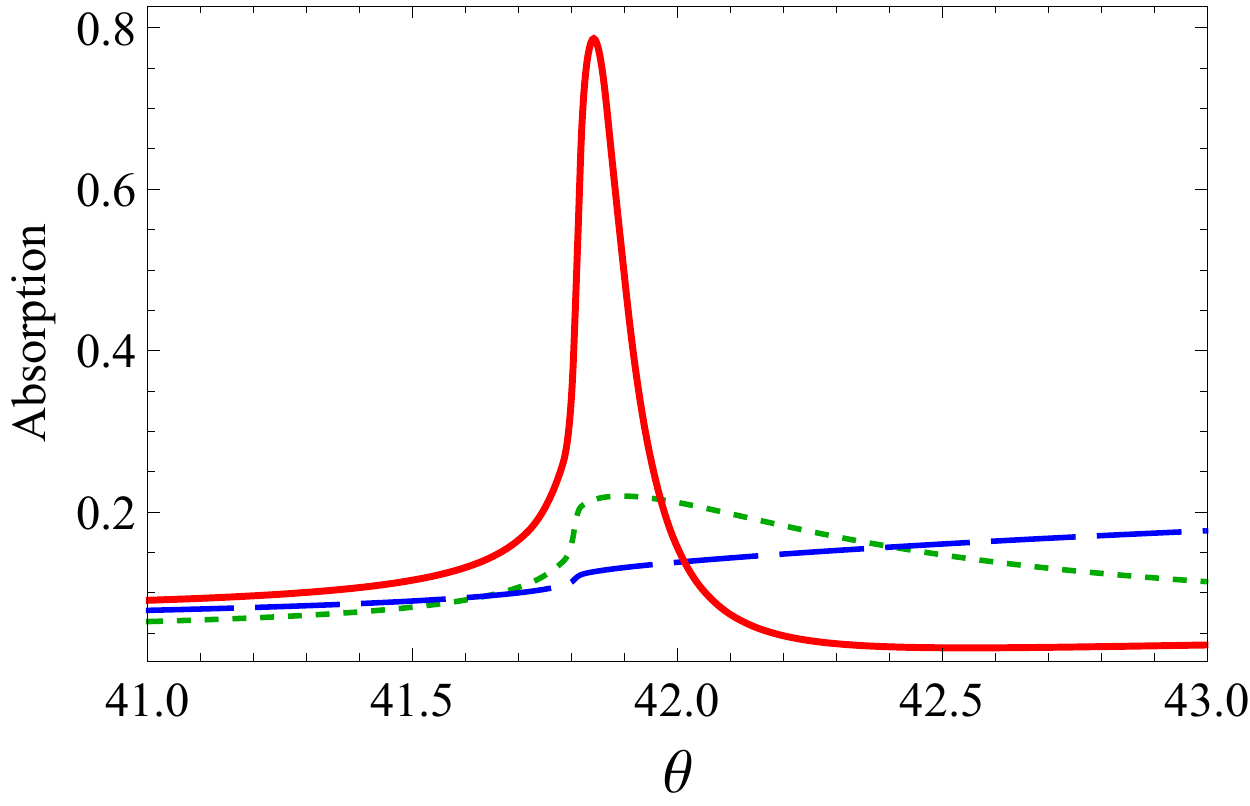}}
	
	\subfigure[]{\includegraphics[width=0.7\columnwidth]{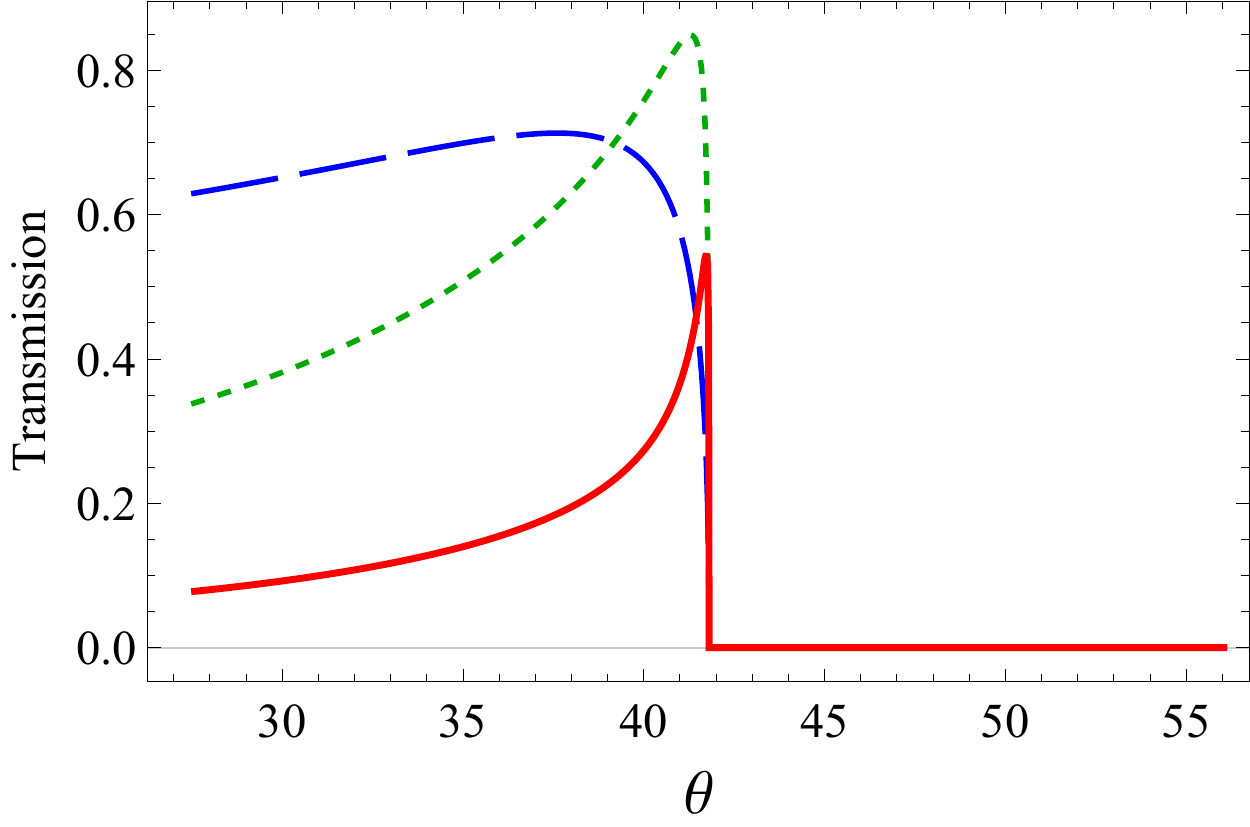}}
	\caption{Field enhancement factor (a), absorption (b), and transmission (c) as a function of the incidence angle $\theta$ (degrees): circles (short dashes), vertical ellipses (long dashes), horizontal ellipses (solid line).}
	\label{f:FEF}
\end{figure}

Formulas for the extinction and scattering are
\begin{equation}\label{extintion}
	C_{ext}=-\frac1{I_0}\int(\mathbf{S}_{ext}\cdot \mathbf{e}_r)\,dA,\quad
	C_{s}=\frac1{I_0}\int(\mathbf{S}_{s}\cdot \mathbf{e}_r)\,dA,
\end{equation}
where $\mathbf{S}_s=[\mathbf{E}_z\times\mathbf{H}_s^*]$, $\mathbf{S}_{ext}=[\mathbf{E}_0\times\mathbf{H}_s^*
+\mathbf{E}_s\times\mathbf{H}_0^*]$ are the Pointing vectors of scattered radiation and the energy flux of interaction between scattered and incident fields, responsible for the extinction. We take the integral over the surface of a cylinder $A$. The absorption coefficient was calculated as the difference $C_a=C_{ext}-C_s$ \cite{BH04,jtpGZ06}. We see that FEF and absorption have close qualitative behavior. The transmission coefficient is calculated in the far-field domain, neglecting all the evanescent modes. The transmission has different angular dependence with a sharp edge at the angle of TIR. 

Figure \ref{f:FEF} shows a comparison of the angular dependencies near the TIR angle of the field enhancement factor (FEF) (a), absorption (b), and transmission coefficients (c) for three versions of cross-sections. Figure \ref{f:FEF}(a) demonstrates the  $\mbox{FEF}=|E/E_0|^2$ in the middle between neighbor cylinders as a function of the incidence angle. Here $E$ is the electric field in the middle of the gap, and $E_0$ is the incident field. A plasmon resonance induces a peak, exceptionally sharp for the horizontal ellipses. We observe the highest peak since the horizontal ellipse has a minimum radius of curvature in the gap. 

As previously mentioned \cite{frumin2020sensitivity}, the resonance of the slit is excited by the refracted into the upper half-space (Fresnel) field. There is only an evanescent wave when $\theta$ exceeds the TIR angle. As a result, the Fresnel field achieves its maximum near the TIR angle; its $x$-component vanishes. The absolute maximum of the intensity is created only by the $y$-component of the Fresnel field. Component $E_x$, and not $E_y$, experiences resonance in the gaps. A natural question arises: how is this possible if the Fresnel $x$-component of the field vanishes. It turns out that the field scattered on the cylinders has a significant $x$-component. Since the cylinders are not limiting subwavelength ($kd=0.9, 0.5, 1.7$ for (i)--(iii) cases), then the scattering differs from the dipole case. The phase difference between neighbor wires originates the additional contribution to the $x$-component. We emphasize that the gaps between the cylinders are subwavelength: $k\Delta <1$, where $\Delta=0.01~\mu$m. 

Absorption, fig. \ref{f:FEF}(b), correlate well with FEF in fig. \ref{f:FEF}(a). It cannot be surprising since the most intensive electric field and the largest damping is achieved precisely in plasmon resonance. The absorption characteristics, to a certain extent, reproduce the angular dependence of FEF. As for the transmittance presented in fig \ref{f:FEF}(c), it vanishes at a greater angle than the TIR. The angular distribution of the horizontal ellipses has a shape similar to a resonance curve. For other cross-sections, we see a substantial broadening. The curve is wider for the vertical ellipses, as well as in FEF and absorption.

\section{Anisotropic layer}

\begin{figure}[h]\centering
	\includegraphics[width=0.4\textwidth]{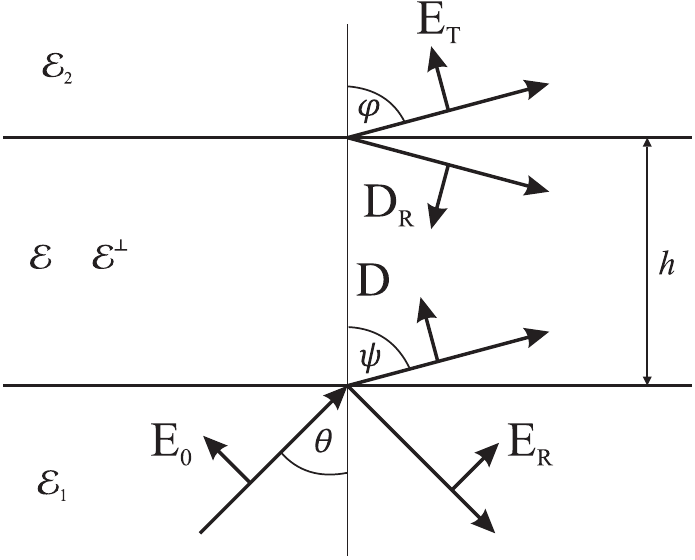}
	\caption{The anisotropic Fresnel's model of a layer  between two half-spaces. Permittivity is $\eps$ for the  parallel electric field and $\eps^\bot$ for transverse field with respect to the boundary.}
	\label{fig:Layer scheme}
\end{figure}

Helmholz equation gives dispersion relations in each part of the area, and boundary conditions match tangential components of electric  and magnetic vectors $\mathbf{E},\mathbf{H}$, respectively:
\begin{eqnarray}\label{tangent_electric}
	- E_0 \cos\theta + E_\mathrm{R} \cos\theta & = & - \frac{D}{\eps} \cos\psi - \frac{D_\mathrm{R}}{\eps} e^{\delta}\cos\psi, \nonumber \\
	- \frac{D}{\eps} e^{\delta}\cos\psi - \frac{D_\mathrm{R}}{\eps} \cos\psi & = & - E_\mathrm{T} \cos\phi \;,\\
\label{perp_electric}
	\eps_1 E_0 \sin\theta + \eps_1 E_\mathrm{R} \sin\theta & = & D \sin\psi - D_\mathrm{R} e^{\delta}\sin\psi, \nonumber\\
	D e^{\delta}\sin\psi - D_\mathrm{R} \sin\psi & = & \eps_2 E_\mathrm{T} \sin\phi \;.
\end{eqnarray}
We consider the scattering of incident wave with electric field $\mathbf{E}_0$ by an isotropic or anisotropic layer of thickness $h$. 

For the isotropic material, we substitute cylinders with dielectric constant $\eps_3$ and surrounding free space with a dielectric or metallic thin film  $\eps$. In anisotropic case the permittivity tensor has two diagonal components $\eps_3^{xx}=\eps,\eps_3^{yy}=\eps^\bot$. The incidence angle is $\theta$, the inside layer refraction angle is $\psi$, and the transmittance angle is $\phi$. Electric fields of reflected and transmitted waves are $\mathbf{E}_\mathrm{R}$ and $\mathbf{E}_\mathrm{T}$, respectively. Vectors of electric displacement $\mathbf{D}$ and $\mathbf{D}_\mathrm{R}$  are orthogonal to their wavevectors inside the layer. 

Figure \ref{fig:Layer scheme}  illustrates the electric field notation: we calculate tangential components by multiplying vector lengths by cosines and sines of corresponding refraction angles.  The imaginary part of cosine or sine turns to zero only in non-absorptive dielectric domains at refraction angles up to TIR. In general, they are complex and should be defined carefully. Furthermore, due to nonzero layer thickness, we take the phase shift into account:
\begin{equation}\label{anisotropic_phase}
	\delta = {i}\frac{2\pi}{\lambda} h\, \sqrt{\eps - \eps_1\frac{\eps}{\eps^\perp} \sin^2\theta }\;.
\end{equation}
The complex angle $\psi$ inside anisotropic layer satisfies relations
\begin{equation}\label{anisotropic_angles}
	\sin\psi = \frac{\sqrt{\eps_1} \sin\theta}{\sqrt{\eps + \eps_1 \left(1 - \frac{\eps}{\eps^\perp}\right) \sin^2\theta}} \;,\quad \cos\psi = \frac{\sqrt{\eps - \eps_1\frac{\eps}{\eps^\perp} \sin^2\theta}}{\sqrt{\eps + \eps_1\left(1 - \frac{\eps}{\eps^\perp}\right) \sin^2\theta}} \;,
\end{equation}
here the cut in complex plane of the square root function goes along the negative real semi-axis. The transmit refraction angle is
\begin{equation}\label{transmit_angles}
	\sin\phi = \frac{\sqrt{\eps_1}}{\sqrt{\eps_2}}\, \sin\theta \;,\quad \cos\phi = \frac{\sqrt{\eps_2 - \eps_1 \sin^2\theta}}{\sqrt{\eps_2}} \;.
\end{equation}

Eq. (\ref{tangent_electric}) and (\ref{perp_electric}) are sufficient to find transmittance
\begin{eqnarray}
	\mathrm{T} = 4\,\frac{\Re(\sqrt{\eps_2}\cos\phi)}{\Re(\sqrt{\eps_1}\cos\theta)} \nonumber\\
	\times\left|\left(\frac{\sqrt{\eps_2}}{\sqrt{\eps_1}} + \frac{\cos\phi}{\cos\theta}\right) \cosh\delta - \left(\frac{\eps}{\eps_1} \frac{\sin\psi}{\sin\theta} \frac{\cos\phi}{\cos\psi} + \frac{\eps_2}{\eps} \frac{\sin\phi}{\sin\psi} \frac{\cos\psi}{\cos\theta}\right) \sinh\delta \right|^{-2} \;,\label{transmit_electric}
\end{eqnarray}
and absolute reflectance
\begin{equation}\label{reflect_electric}
	\mathrm{R} = \left|\frac{\left(\frac{\sqrt{\eps_2}}{\sqrt{\eps_1}} - \frac{\cos\phi}{\cos\theta}\right) \cosh\delta - \left(\frac{\eps}{\eps_1} \frac{\sin\psi}{\sin\theta} \frac{\cos\phi}{\cos\psi} - \frac{\eps_2}{\eps} \frac{\sin\phi}{\sin\psi} \frac{\cos\psi}{\cos\theta}\right) \sinh\delta} {\left(\frac{\sqrt{\eps_2}}{\sqrt{\eps_1}} + \frac{\cos\phi}{\cos\theta}\right) \cosh\delta - \left(\frac{\eps}{\eps_1} \frac{\sin\psi}{\sin\theta} \frac{\cos\phi}{\cos\psi} + \frac{\eps_2}{\eps} \frac{\sin\phi}{\sin\psi} \frac{\cos\psi}{\cos\theta}\right) \sinh\delta}\right|^2 \;.
\end{equation}
These coefficients mean the ratio of transmitted through the layer and reflected energy flux to incident one. So the sum of transmittance $\mathrm{T}$, reflectance $\mathrm{R}$ and absorption $\mathrm{A}$ add up to one. Then to get the absorption coefficient we apply relation $\mathrm{A} = 1 - \mathrm{T} - \mathrm{R}$. Here the absorption is specific, i.e.,  normalized by period $d$ and $\cos\theta$. If phase shift $\delta$ tends to zero Eq.  (\ref{transmit_electric}) and (\ref{reflect_electric}) become simpler and turn to Fresnel's ones for two half-spaces with common boundary plane. If anisotropic layer becomes isotropic $\eps = \eps^\bot$, then the transmittance (\ref{transmit_electric}) coincides with found one for magnetic field \cite{K66,R88}. Note that the formulas for isotropic medium are applicable also for reflectometry sensor based on metallic-dielectric structure \cite{terent2015,Goldina16}. 

\begin{figure}\centering
	\subfigure[]{\includegraphics[width=0.8\columnwidth]{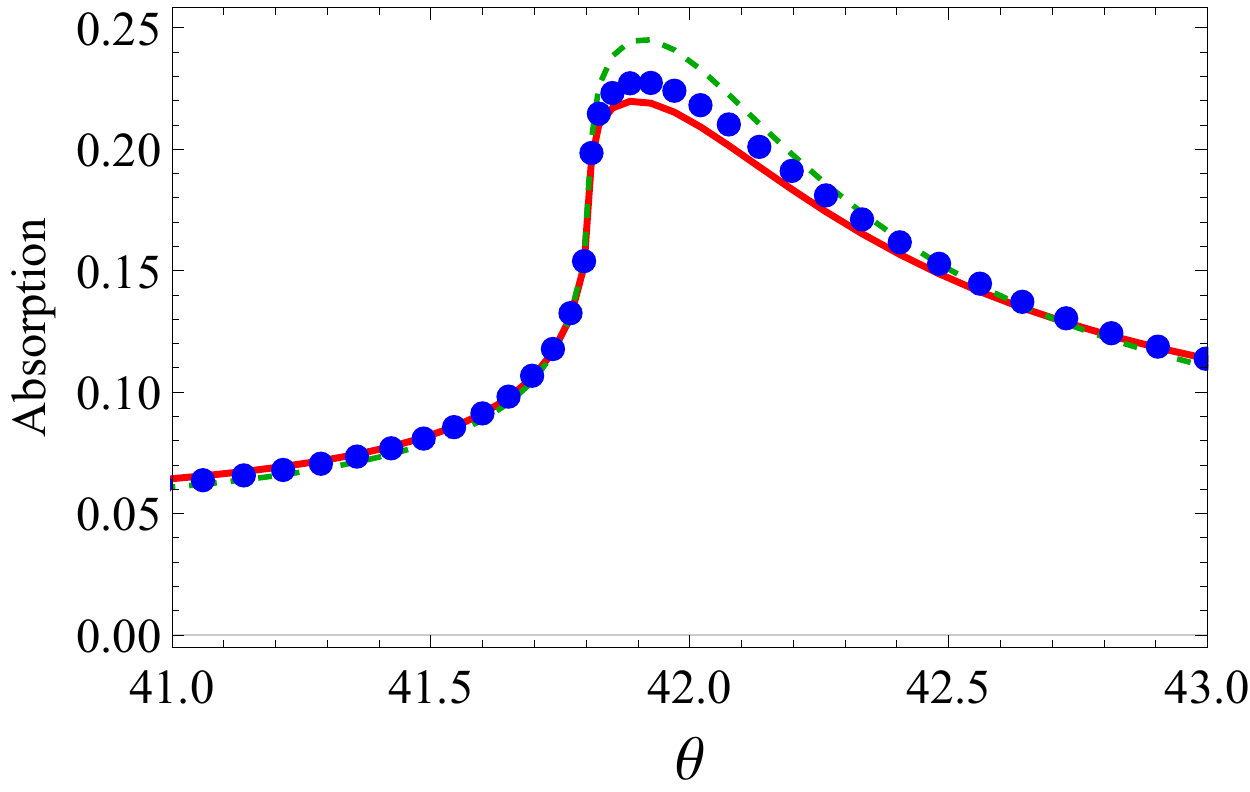}}
	\subfigure[]{\includegraphics[width=0.8\columnwidth]{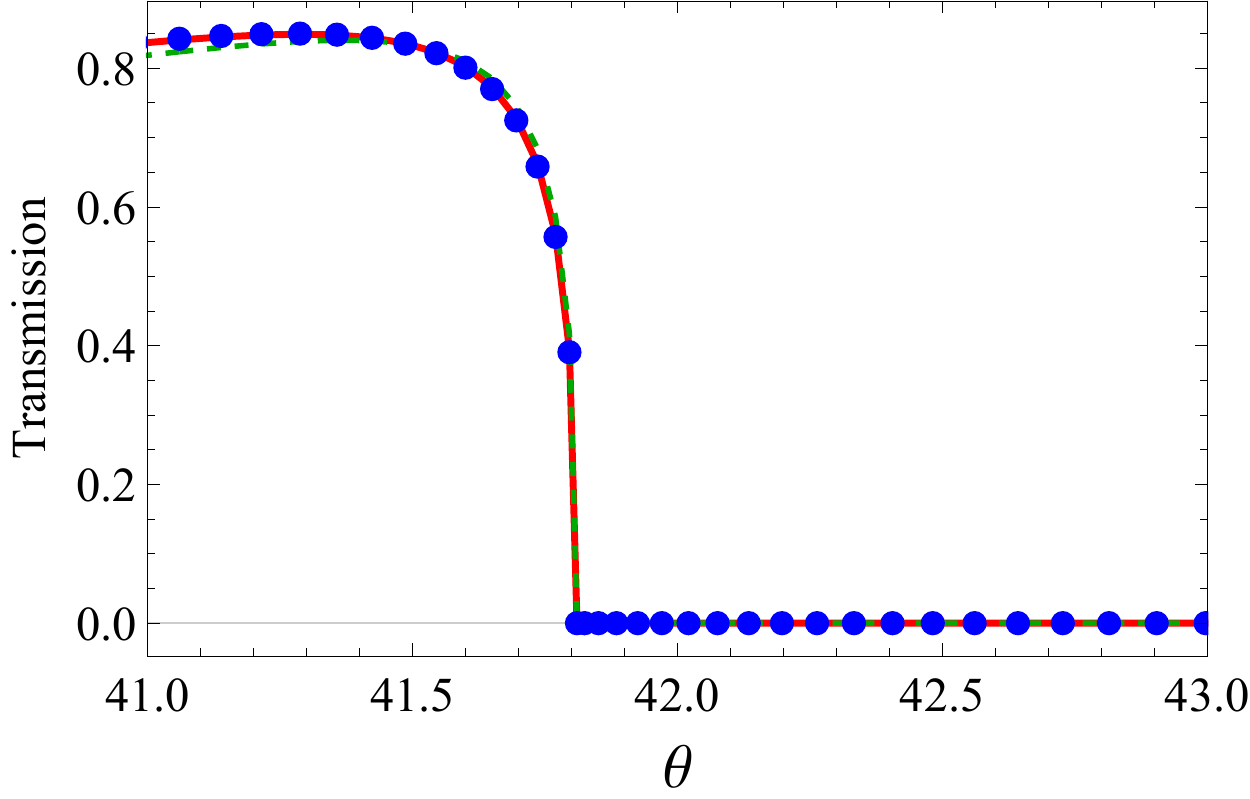}}
	\caption{Absorption (a) and transmission (b) of circular cylinders as a function of the incidence angle $\theta$ (degrees): BEM calculation (solid line), isotropic layer (dashes), non-isotropic medium (circles).}
	\label{f:circular-fitting}
\end{figure}

\begin{figure}\centering
	\subfigure[]{\includegraphics[width=0.8\columnwidth]{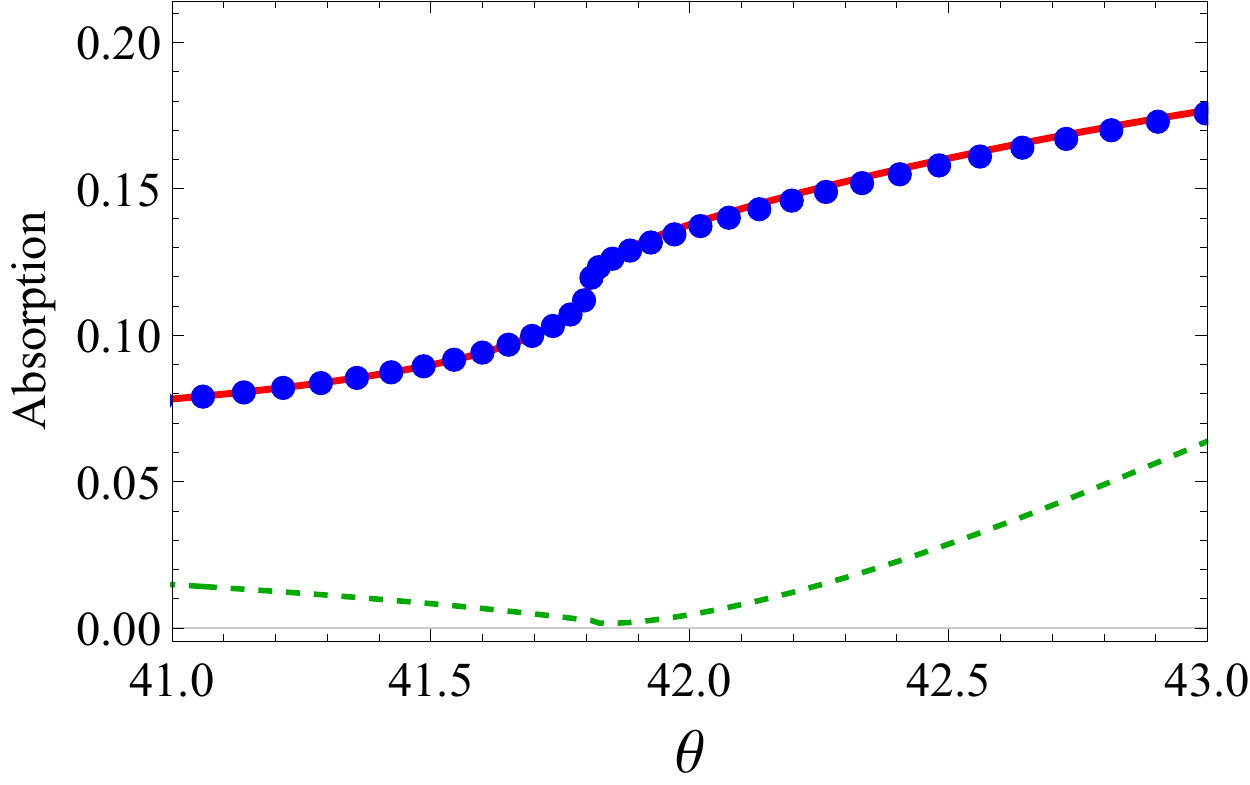}}
	\subfigure[]{\includegraphics[width=0.8\columnwidth]{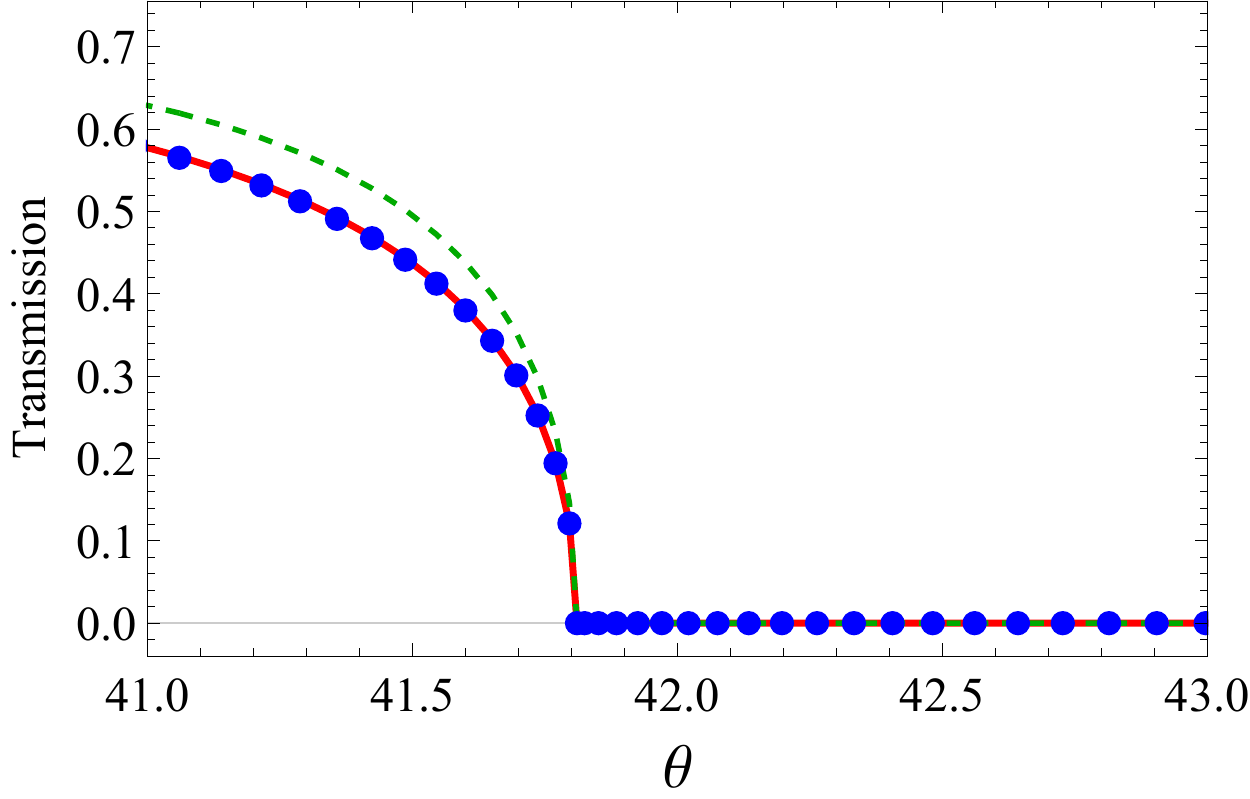}}
	\caption{The same as in fig. \protect\ref{f:circular-fitting}, but for elliptic cylinders with vertical orientation of large axis.}
	\label{f:vertical-fitting}
\end{figure}

\begin{figure}\centering
	\subfigure[]{\includegraphics[width=0.8\columnwidth]{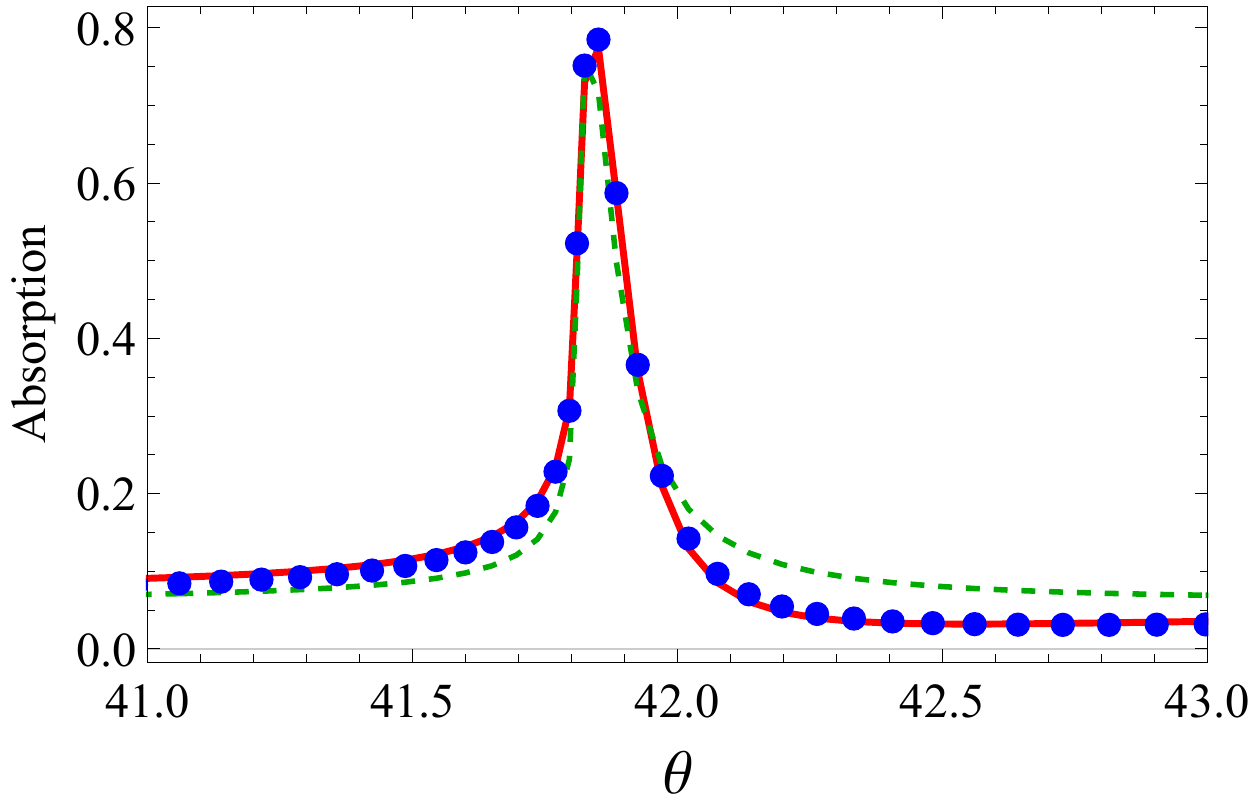}}
	\subfigure[]{\includegraphics[width=0.8\columnwidth]{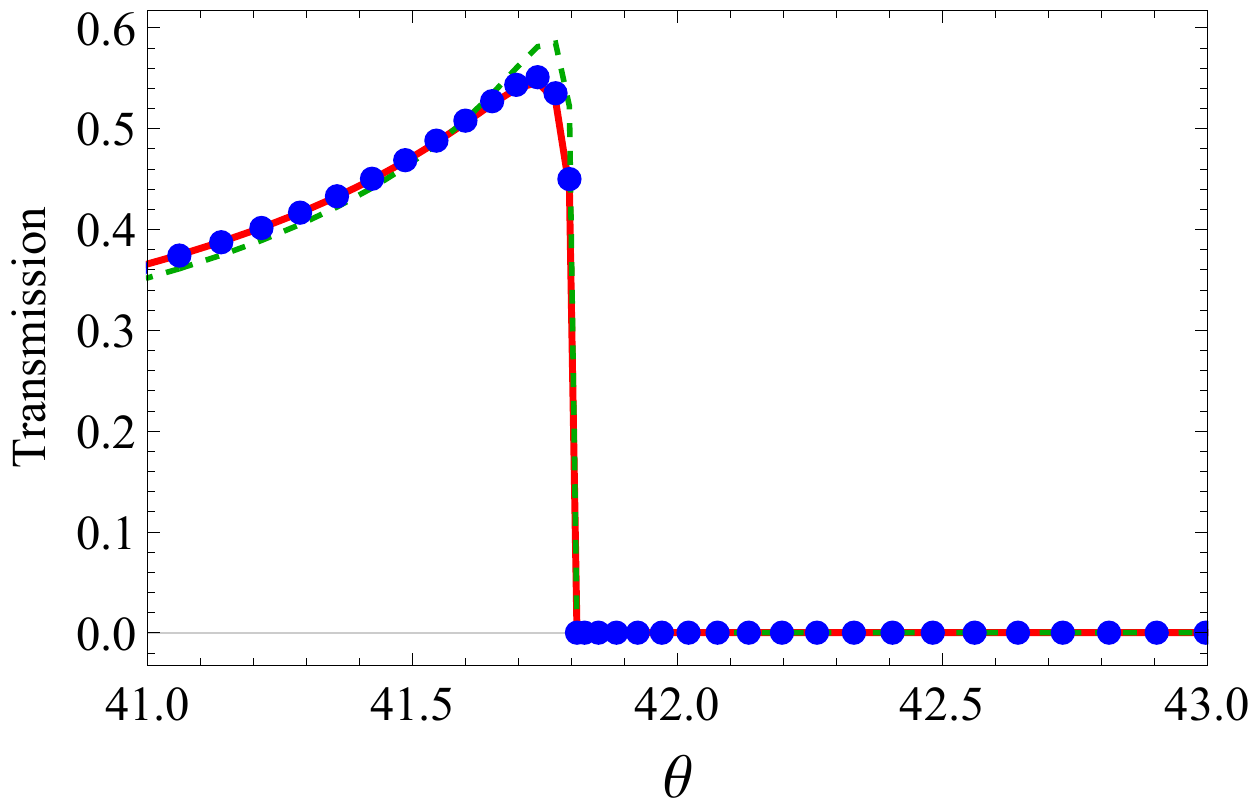}}
	\caption{The same as in fig. \protect\ref{f:circular-fitting}, but for elliptic cylinders with horizontal orientation of large axis.}
	\label{f:horizontal-fitting}
\end{figure}

Let us plot the angular distribution within the isotropic and anisotropic layer models and compare them with numerical results. We fit the model parameters (thickness and permittivity) to minimize the offsets by the least-squares method. The functional includes both the residues of transmission and specific absorption coefficients with equal weights:
\begin{equation}\label{Distortion_functional}
	F = \sum_{n = 1}^N \left|T(\theta_i)-T_i\right|^2 + \sum_{n = 1}^N \left|A(\theta_i)-A_i\right|^2\;,
\end{equation}
here $T(\theta),A(\theta)$ are the model transmittance or specific absorption depending on refraction angle in glass, $T_i,A_i$ are the transmittance or absorption calculated by BEM numerically at the incidence angle $\theta_i$ in glass. Number $N=201$ corresponds to numerical samples of incidence angle  glass in our calculations. The best fit parameters of the least-squares procedure are gathered in table \ref{table:Summary table}.

How accurate the fit for absorption and transmission is, one can see from 
fig. \ref{f:circular-fitting}, \ref{f:vertical-fitting}, \ref{f:horizontal-fitting}. The isotropic model describes the BEM curves qualitatively. When the real part of permittivity occurs negative (the third line and third column of the table), the fitting is good in transmission but loses its accuracy in absorption. Otherwise, the curve has a huge mismatch in transmission and better fitting in absorption, though the distortion function remains large. For example, it is negative for the isotropic (scalar) model of vertical ellipses,  fig. \ref{f:vertical-fitting}(a). Meanwhile, the anisotropic model describes all the curves quantitatively with high accuracy. It becomes evident from the bottom line in table \ref{table:Summary table}. The better fitting by the anisotropic model is at least due to more fitting parameters (5 vs. 3).

The anisotropy turns out to be sufficiently high in this structure. Different signs of the real part in the second and third lines from the bottom indicate that the media behaves as hyperbolic metamaterials, i.e., a medium with the hyperbolic dispersion relation \cite{poddubny2013hyperbolic}.  One could explain this observation within the concatenated capacitor model in the circuit diagram. For circular cylinders (nanowires) at a small fill-fraction of the metal compared to dielectric media in a unit cell, the Maxwell --- Garnett approach is exploited to calculate the dielectric tensor components for metamaterials \cite{cortes2012quantum}.

\begin{table}\centering
	\caption{Resulting  parameters in isotropic and anisotropic layer models.}
	\label{table:Summary table}
\begin{tabular}{|p{4cm}|p{2.4cm}|p{2.4cm}|p{2.4cm}| }
\hline
	Parameter & Circle & Vertical ellipse   & Horizontal ellipse  \\
\hline
	Isotropic/anisotropic layer thickness, nm $(h)$ & {55/40} & {9.2/103} & {24/19} \\
\hline
	Isotropic permittivity $(\eps)$ & $20.7+0.7\rm{i}$ & {$-27.6+1.5\rm{i}$} & {$93.7+8.6{i}$} \\
	\hline
	Longitudinal permittivity $(\eps)$ & $27.0+1.0{i}$ & $-0.95+0.08{i}$ & $\;117+19.1{i}$ \\
\hline
Transversal permittivity $(\eps^\bot)$ &
$-2.55+0.0{i}$ &$1.34+0.04{i}$ &
 $-2.96-0.55{i}$ \\
\hline
Isotropic/anisotropic layer distortion functional $(F \cdot 10^3)$& {65.3/4.7} & {705/0.67} & {149/11.5}\\
\hline
\end{tabular}
\end{table}

\section{Conclusions}

To study the angular dependence of scattering parameters near the TIR angle, we perform numerical calculations by the BEM. 
We consider circular and elliptic cylinders with vertical and horizontal large-axis orientations and treat the changes in absorption, transmittance, and the field enhancement factor. We examine two possible directions of the large ellipse axis: horizontal and vertical. The plasmonic resonance is sharper for horizontal ellipses, where the curvature radius at the slit is the least.

We compare the calculated absorption and transmission coefficient with the `sandwich' model. In the model, we place a medium with some effective average permittivity and fixed thickness between two dielectric half-spaces instead of the parallel wires. This model allows the analytical expressions of transmission and absorption coefficients. The least-square minimization shows that both models reproduce well the behavior of curves. Moreover, the curves practically coincide with BEM computation when one chooses the model of the anisotropic layer. Therefore, the anisotropic model is more appropriate to describe artificial nanostructured media known as metamaterials.

\section*{Acknowledgments}
We are grateful to O. V. Belai  for helpful discussions.
This work is supported by the Russian Foundation of Basic Research, grant \#~20-02-00211.


\end{document}